\documentclass[12pt]{article}
\usepackage{latexsym}
\usepackage{graphicx}
\usepackage{times}

\begin{document}
\title{Locality of quantum entanglement}
\author{Wang Guowen\\College of Physics, Peking University, Beijing, China}
\date{December 22, 2005}
\maketitle

\begin{abstract}
This article presents a local realistic interpretation of quantum
entanglement. The entanglement is explained as innate interference
between the non-empty state associated with the peaked piece of
one particle and the empty states associated with the non-peaked
pieces of the others of entangled particles, which inseparably
join together. The correlation of the results of measurements on
the ensemble of composite entangled systems is related to this
kind of interference. Consequently, there is no nonlocal influence
between entangled particles in measurements. Particularly, this
explanation thus rules out the possibility of quantum
teleportation which is nowadays considered as one of cornerstones
of quantum information processing. Besides, likewise,
communication and computation schemes based on alleged spooky
action at a distance are unlikely to be promising.
\end{abstract}

\section{Introduction}
Recently, the application of quantum mechanics enters the field of
information science and technology. Some current theories of
quantum information seem to be dependent on interpretation of
quantum mechanics. The local and nonlocal interpretations of
quantum mechanics lead to different consequences. For example, the
latter leads to the idea of quantum teleportation. Quantum
teleportation refers to the scenario that a quantum state has
disappeared from a location and repeated instantaneously in a
distant region. Bennett \emph{et al.} in 1993 claimed: $``$An
unknown quantum state $|\phi\rangle$ can be disassembled into,
then later reconstructed from, purely classical information and
purely nonclassical Einstein-Podolsky-Rosen (EPR) correlation.$"$
[1] However Einstein \emph{et al.} (EPR) denied existence of any
nonlocal influences between spacelike separated particles and
concluded that the quantum-mechanical description is incomplete.
At the beginning of their paper [2], they stated: $``$Any serious
consideration of a physical theory must take into account the
distinction between the objective reality, which is independent of
any theory, and the physical concepts with which the theory
operates. These concepts are intended to correspond with the
objective reality, and by means of these concepts we picture this
reality to ourselves.$"$ And at its end they concluded: $``$While
we have thus shown that the wave function does not provide a
complete description of the physical reality, we left open the
question whether or not such a description exists. We believe,
however, that such a theory is possible.$"$ They did not mention
any types of hidden variables in their paper. Dirac said: $``$I
think that it is quite likely that at some future time we may get
an improved quantum mechanics in which there will be a return to
determinism and which will, therefore, justify the Einstein point
of view.$"$ [3] Unfortunately, the EPR argument has been
considered as a failure by the majority of physicists. In favor of
Einstein's local realism the present author will give a local
realistic interpretation of quantum entanglement and shows the
impossibility of quantum teleportation.

\section{Locality of quantum entanglement}
The explanation of entanglement of a composite quantum system is
fundamentally related to the description of a single system. In
Ref.[4] the author describes a free particle as linear
non-spreading wave packet in the framework of special relativity
and logically explains quantum interference experiments. The wave
packet consists of countless Fourier components among which there
is one called characteristic component which frequency and wave
vector are related to the energy and momentum of the particle,
respectively. The Schr\"{o}dinger wave packet or light pulse
consists of such characteristic components which are related to
different energies and momenta and thus spreads in principle. The
non-spreading wave packet contains a peak where the phases of all
the components are same and its off-peak part where the
distribution of their phases makes the resulting amplitude nearly
infinitesimal. Diffraction and interference phenomena demonstrate
that cutting a piece away from the off-peak part or recombining it
will change the path of the peak and/or the energy of the
particle. Different from classical reality, a quantum reality such
as photon, electron or atom, has to be considered to contain a
peak and its off-peak part which plays an essential role in
quantum interference and is the root of quantum weirdness. The
concept of a point-like classical particle reflects the ignorance
of the existence of the off-peak part.

In a composite system containing two or more particles in a
superposition state, according to Schr\"{o}dinger [5], there
exists entanglement. From the concept of the non-spreading wave
packet as being quantum reality, the entanglement of a pair of
particles can be explained as innate interference between the
peaked piece of one particle and the non-peaked piece of the other
of the pair, which inseparably join together. Yet the two peaked
pieces keep mutual independence if the two peaks are spacelike
separated. That is, quantum entanglement is local and cannot be
regarded as $``$spooky action at a distance$"$ (Einstein's words
[6], March 1947).

For example, we consider a pair of entangled particles described
by the unnormalized wave function
\begin{equation} \label{eq1}
\Psi_{p_0}(x_1,x_2)=\frac{1}{\sqrt{2}} (e^{-ip_0x_1/\hbar}\cdot
e^{ip_0x_2/\hbar}+e^{-ip_0x_2/\hbar}\cdot e^{ip_0x_1/\hbar})
\end{equation}
which is interpreted as a probability amplitude. In order to make
quantum-mechanical description formally more complete, let's use
the symbol $\#...\#$ to denote the function which is the
characteristic component associated with the non-peaked piece of a
quantum system. We will call the state indicated by this symbol
$``$empty state$"$ which carries information on the non-peaked
piece. So the real state of the pair of particles should be
assumed to be one of the following two:
\begin{equation}
\label{eq2}
\Psi_{p_0,-p_0}(x_1,x_2)=\frac{1}{\sqrt{2}}(e^{-ip_0x_1/\hbar}\cdot
e^{ip_0x_2/\hbar}+\# e^{-ip_0x_2/\hbar} \# \cdot \#
e^{ip_0x_1/\hbar} \#)
\end{equation}
\begin{equation}
\label{eq3} \Psi_{-p_0,p_0}(x_1,x_2)=\frac{1}{\sqrt{2}}(\#
e^{-ip_0x_1/\hbar} \# \cdot \# e^{ip_0x_2/\hbar}
\#+e^{-ip_0x_2/\hbar}\cdot e^{ip _0x_1/\hbar})
\end{equation}
These equations show that there is innate interference between the
non-empty state associated with the peaked piece of one particle
and the empty state associated with the non-peaked piece of the
other of the pair, which inseparably join together. Consequently,
the correlation of results of measurements on the ensemble of the
pairs is related to this kind of interference. So the correlation
characterizing quantum entanglement is local.

Now we consider the case in the position representation where a
pair of particles is in the entangled state
\begin{equation} \label{eq4}
\Psi_{x_0}(x_1,x_2)=\frac{1}{\sqrt{2}}
(\delta(x_0-x_1)\cdot\delta(x_0+x_2)+\delta(x_0-x_2)\cdot\delta(x_0+x_1))
\end{equation}
The real state of the pair is one of the following two:
\begin{equation} \label{eq5}
\Psi_{x_0,-x_0}(x_1,x_2)=\frac{1}{\sqrt{2}}
(\delta(x_0-x_1)\cdot\delta(x_0+x_2)+\#\delta(x_0-x_2)\#\cdot
\#\delta(x_0+x_1)\#)
\end{equation}
\begin{equation} \label{eq6}
\Psi_{-x_0,x_0}(x_1,x_2)=\frac{1}{\sqrt{2}}
(\#\delta(x_0-x_1)\#\cdot\#\delta(x_0+x_2)\#+\delta(x_0-x_2)\cdot\delta(x_0+x_1))
\end{equation}
Eq.4 implies that after the position measurement on both
particles, particle 1 and 2 will have been found simultaneously at
$x_1=x_0$ and $x_2=-x_0$ or at $x_1=-x_0$ and $x_2=x_0$, or
implies equivalently in the statistical sense that particle 1 of a
pair expressed by Eq.5 and particle 2 of another pair expressed by
Eq.6 in their pair ensemble will have been found simultaneously at
$x_1=x_2=x_0$ with equal probability after the position
measurement at that place.

Furthermore, in order to discuss measurement effects on the
entangled pair we consider the case where the two particles of the
entangled pair described by Eq.1 are spacelike separated. If only
particle 1 of the pair has been measured and found at $x_1=x_0$,
the pair will have jumped into the real state
\begin{equation} \label{eq7}
\Psi_{x_0,-p_0}(x_1,x_2)=\frac{1}{\sqrt{2}}(\delta(x_0-x_1)\cdot
e^{ip_0x_2/\hbar}+\#\delta(x_0-x_2)\#\cdot\#e^{ip_0x_1/\hbar}\#)
\end{equation}
or if only particle 2 has been measured and found at $x_2=-x_0$,
the pair will have jumped into the real state
\begin{equation} \label{eq8}
\Psi_{p_0,-x_0}(x_1,x_2)=\frac{1}{\sqrt{2}}
(e^{-ip_0x_1/\hbar} \cdot \delta(x_0+x_2)+\#e^{-ip_0x_2
/\hbar}\#\cdot \#\delta(x_0+x_1)\#)
\end{equation}
These equations show that there is no nonlocal influence between
the two entangled particles in the measurement. Let's explain it
in more details. Assuming that Alice locates on the positive
half-axis of \emph{x} and Bob on the negative one, if she has
found particle 1  at $x_1=x_0$ after her measurement using a tool
such as a clamp equivalent to an extremely narrow and deep
potential well, the state $e^{-ip_0x_1/\hbar}$ of the particle
will have evolved into $\delta (x_1-x_0)$ which is an eigenstate
of the position operator X, while the state $e^{ip_0x_2/\hbar}$ of
particle 2 remains intact since the two associated peaks are
spacelike separated. On the other hand, the empty state
$\#e^{-ip_0x_2/\hbar}\#$ of particle 2 will also have evolved
simultaneously into the empty state $\#\delta (x_2-x_0)\#$ because
of action of the clamp, while the empty state
$\#e^{ip_0x_1/\hbar}\#$ of particle 1 remains intact. It is
similar for the case where Bob performs a position measurement on
particle 3 in the same way. So quantum entanglement has no
nonlocality feature. Thus the Bell inequalities [7] are not the
touchstones for judging whether quantum mechanics is local or
nonlocal. It is a mistake that the Bell inequality violation is
interpreted as evidence for nonlocal influences in the measurement
on entangled particles. We see that containing a return to
determinism by supplementing the non-peaked piece of a system as a
new kind of quantum reality, the improved quantum mechanics yet
completely tallies predictions for correlation measurement led by
quantum entanglement. We are aware that quantum mechanics does not
explicitly require and yet not implicitly rule out this kind of
supplementary reality. However, the proper interpretation of
quantum mechanics exactly needs the supplement as a basis.

\section{Impossibility of quantum teleportation}
Now we are in position to discuss whether or not quantum
teleportation is possible. According Bennett \emph{et al.} [1], if
Alice and Bob share the entangled state of a pair of particles
with spin-$\frac{\tiny{1}}{\tiny{2}}$
\begin{equation} \label{eq9}
|\Psi^{-}_{23}\rangle=\frac{1}{\sqrt{2}}(|\uparrow _2\rangle
|\downarrow _3\rangle-|\uparrow _3\rangle|\downarrow _2\rangle)
\end{equation}
Alice can send Bob an unknown arbitrary state
\begin{equation} \label{eq10}
|\phi_1\rangle=\alpha|\uparrow _1\rangle +\beta|\downarrow
_1\rangle, \mbox{ }|\alpha|^2+|\beta|^2=1
\end{equation}
by making use of the following direct product state
\begin{eqnarray}\label{eq11}
|\Psi_{123}\rangle=|\phi_1\rangle|\Psi^{-}_{23}\rangle
=\frac{1}{2}[|\Psi^{+}_{12}\rangle(-\alpha|\uparrow
_3\rangle+\beta|\downarrow _3\rangle)+
|\Psi^{-}_{12}\rangle(-\alpha|\uparrow _3\rangle-\beta|\downarrow
_3\rangle)\nonumber\\
+|\Phi^{+}_{12}\rangle(-\beta|\uparrow _3\rangle+\alpha|\downarrow
_3\rangle)+|\Phi^{-}_{12}\rangle(\beta|\uparrow
_3\rangle+\alpha|\downarrow _3\rangle)]\nonumber\\
=\frac{1}{2}[|\Psi^{+}_{12}\rangle M_{11}|\phi_3\rangle+
|\Psi^{-}_{12}\rangle M_{00}|\phi_3\rangle +|\Phi^{+}_{12}\rangle
M_{10}|\phi_3\rangle+|\Phi^{-}_{12}\rangle
M_{01}|\phi_3\rangle],\nonumber\\
|\phi_3\rangle=\alpha|\uparrow _3\rangle +\beta|\downarrow
_3\rangle
\end{eqnarray}
where the set
\begin{equation} \label{eq12}
|\Psi^{\pm}_{12}\rangle=\frac{1}{\sqrt{2}}(|\uparrow _1\rangle
|\downarrow _2\rangle\pm|\uparrow _2\rangle|\downarrow _1\rangle),
\mbox{ }|\Phi^{\pm}_{12}\rangle=\frac{1}{\sqrt{2}}(|\uparrow
_1\rangle |\uparrow _2\rangle\pm|\downarrow _2\rangle|\downarrow
_1\rangle)
\end{equation}
is the so-called bell basis, and
\begin{equation} \label{eq13}
M_{00}=-I, \mbox{} M_{01}=\sigma_x, \mbox{} M_{10}=-i\sigma_y,
\mbox{} M_{11}=-\sigma_z
\end{equation}
in which $I$ is the identity operator and $\sigma_x, \sigma_y$,
$\sigma_z$ are the Pauli operators:
\begin{eqnarray}\label{eq14}
I=\left(%
\begin{array}{cc}
   1 & 0 \\
  0 & 1 \\
\end{array}%
\right),\mbox{ }
\sigma_x=\left(%
\begin{array}{cc}
   0 & 1 \\
  1 & 0 \\
\end{array}%
\right), \mbox{ }
\sigma_y=\left(%
\begin{array}{cc}
   0 & -i \\
  i & 0 \\
\end{array}%
\right), \mbox{ }
\sigma_z=\left(%
\begin{array}{cc}
   1 & 0 \\
  0 & -1 \\
\end{array}%
\right)
\end{eqnarray}
With the generally accepted viewpoint, after Alice's  Bell-basis
measurement on particle 1 and 2 , Bob's particle 3 will have
collapsed into a state $M_{i'j'}|\phi_3\rangle$, namely one of the
four states $M_{ij}|\phi_3\rangle, i,j=0, 1$. Consequently, in
their opinion, if Alice communicates through a classical channel
to Bob 2-bit information $(i',j')$ representing the result of the
measurement, Bob $``$can construct an accurate replica
$|\phi_1\rangle$$"$ by using the inverse operation $M^{-1}_{i'j'}$
on his $M_{i'j'}|\phi_3\rangle$. So the state is teleported in
such a way from Alice to Bob with infinite precision (i.e.
infinite amount of data) only at a cost of 2-bit information. Yet
it is said that such a cheap transmission of information is
warranted by quantum mechanics principle. Quantum teleportation is
nowadays considered as one of cornerstones of quantum information
processing.

However, according to the above local realistic interpretation,
before Alice's Bell-basis measurement, particle 3 in Bob's hands
is in one of the following two real states:
\begin{equation} \label{eq15}
|\Psi^{-}_{23}\rangle=\frac{1}{\sqrt{2}}(|\uparrow _2\rangle
|\downarrow _3\rangle-\#|\uparrow _3\rangle \#\#|\downarrow
_2\rangle \#)
\end{equation}
\begin{equation} \label{eq16}
|\Psi^{-}_{23}\rangle=\frac{1}{\sqrt{2}}(\#|\uparrow _2\rangle\#\#
|\downarrow _3\rangle \#-|\uparrow _3\rangle |\downarrow
_2\rangle)
\end{equation}
After her measurement,  on her side the empty state $\#|\uparrow
_3\rangle \#$ or $\#|\downarrow _3\rangle \#$ of particle 3 will
have evolved into one of the four empty states
$M_{ij}(\alpha\#|\uparrow _3\rangle \#+\beta\#|\downarrow
_3\rangle\#), i,j=0, 1$, while on Bob's side the non-empty state
$|\uparrow _3\rangle$ or $|\downarrow _3\rangle $ of particle 3
remains intact. In reality, her measurement only influences the
state $|\phi_1\rangle$ of particle 1 and both the inseparable
interfering states, namely the non-empty state of particle 2 and
the empty state of particle 3 on the side of herself. Thus the
local realistic interpretation rules out the possibility of
quantum teleportation. We argue that the idea of quantum
teleportation is based on the misunderstanding of quantum
mechanics and is invalid. Besides, likewise, communication and
computation schemes based on alleged spooky action at a distance
are unlikely to be promising. Recently, a lot of researchers have
spent much effort to do study of realization of quantum
teleportation [8-15]. Sorry, to the best of the author's
knowledge, until now there seems to be, however, no experiment
that deserves to be called accomplishing successful quantum
teleportation and checks definitely its possibility.

\section{Conclusion}
This article presents a local realistic interpretation of quantum
entanglement. The entanglement is explained as innate interference
between the non-empty state associated with the peaked piece of
one particle and the empty states associated with the non-peaked
pieces of the others of entangled particles, which inseparably
join together. The correlation of the results of measurements on
the ensemble of composite entangled systems is related to this
kind of interference. Consequently, there is no nonlocal influence
between entangled particles in measurements. Particularly, this
explanation thus rules out the possibility of quantum
teleportation which is nowadays considered as one of cornerstones
of quantum information processing. Besides, likewise,
communication and computation schemes based on alleged spooky
action at a distance are unlikely to be promising.

\end{document}